\documentclass{chjaa}                  % preprint, the final version for publication
\usepackage{graphicx,times}             %for PS/EPS graphics inclusion, new
\newcommand{\ecs}{erg cm$^{-2}$ s$^{-1}$}
\newcommand{\es}{erg s$^{-1}$}
\input{epsf.sty}                        %for PS/EPS graphics inclusion, old
\input{psfig.sty}                       %for PS/EPS graphics inclusion, old
\begin{document}

   \title{Are Seyfert 2 Galaxies without Polarized Broad Emission Lines 
More Obscured?
%\footnotetext{$*$ Supported by the National Natural Science Foundation of China.}
}
%   \subtitle{I. Place Your Subtitle Here}

   \volnopage{Vol.0 (200x) No.0, 000--000}      %%preserved for Editor. DOn't remove!
   \setcounter{page}{1}           %%starting page, preserved for Editor. DOn't remove!

   \author{Xin-Wen Shu
      \inst{1,2}\mailto{}
%% Please move "\mailto{}" to the corresponding author of the paper.
   \and Jun-Xian Wang
      \inst{1,2}
   \and Peng Jiang
      \inst{1,2}
      }

   \institute{%National Astronomical Observatories, Chinese Academy of Sciences,
             %Beijing 100012, China
              Center for Astrophysics, University of Science and 
Technology of China (USTC), Hefei, Anhui 230026, China\\
             \email{xwshu@mail.ustc.edu.cn}
%% offprint requests will be sent. Note to pair \mailto{} with \email{}
        \and
             Joint Institute for Galaxy and Cosmology, USTC and 
Shanghai Astronomical Observatory, Chinese Academy of Sciences, 
Hefei, Anhui 230026, China\\
        %\and
         %    Astronomisches Institut der Universit\"{a}t Wien, T\"{u}rkenschanzstr. 17,
          %   A-1180 Wien, Austria\\
          }

   \date{Received~~2001 month day; accepted~~2001~~month day}

   \abstract{
   The new $XMM-Newton$ 
data of seven Seyfert 2 galaxies with optical
spectropolarimetric observations are presented.
 The analysis of 0.5 -- 10 keV spectra shows 
that all four Seyfert 2 galaxies with polarized
broad lines (PBLs) are 
absorbed with $N_{\rm H}<10^{24}$ cm$^{-2}$,
 while two of three Seyfert 2 galaxies without PBLs have evidence suggesting
Compton-thick obscuration, supporting the conclusion that Seyfert 2 galaxies
without PBLs are more obscured than those with PBLs. Adding the measured
obscuration indicators ($N_{\rm H}$,
$T$ ratio, and
Fe K$\alpha$ line EW) of six luminous AGNs to our previous sample
improves the significance level of the difference in absorption 
from 92.3\% to 96.3\%
for $N_{\rm H}$, 99.1\% to 99.4\% for $T$ ratio, and 95.3\%
to 97.4\% for Fe K$\alpha$ line EW. The present results support and enhance the
suggestions that
the absence of PBLs in Seyfert 2 galaxies can be
explained by larger viewing angle of line of sight to the
putative dusty torus, which lead to the obscuration of broad-line
scattering screen, as expected by the unification model.
   }

   \authorrunning{X.-W. Shu, J.-X. Wang \& P. Jiang }            %author_head in even pages
   \titlerunning{Are Seyfert 2 Galaxies without PBLs More Obscured?}  % title_head in odd pages

   \maketitle
%% The author head (on even pages) and the title head (on odd pages) will be
%% automatically extracted from \author{} and \title{}. Whenever the title is too long,
%% you will be asked to supply a shorter one by inserting either \authorrunning{} or
%% \titlerunning{} before \maketitle. Anyway, you can specify your own heads in advance.
%%
%%
%% Note: In the following text body of your manuscript, please note several differences from
%%       other major journals:
%% (1) \subsection{Please Capitalize the First Letter of Each Notional Word in Subsection Title}
%% (2) Please Capitalize the First Letter of Each Notional Word in all tables' captions

%
%________________________________________________ sections below
%
\section{Introduction}           %% first-level sections will be auto-capitalized
\label{sect:intro}

The Seyfert unification model postulates that Seyfert 1 and 2 galaxies
 (hereafter, Sy1s and Sy2s) are intrinsically the same and
 the existence of an optical thick region (the "torus") obscures
 the broad-line region (BLR) in Sy2s (Antonucci 1993). The orientation of this torus relative to our
 line of sight then determines whether the galaxy is classified as a
Seyfert 1 or 2 galaxy.
The most important observational evidence for this orientation-based
unification model is the detection of polarized broad emission lines 
(hereafter PBLs)
in Sy2s in optical spectropolarimetry (Antonucci \& Miller 1985;
Miller \& Goodrich 1990; Young et al. 1996; Heisler et al. 1997;
Moran et al. 2000; Lumsden et al. 2001, 2004; Tran 2001).
Additional evidence for this simple unification model comes from X-ray
studies which have demonstrated that many Sy2s show heavy absorption
along the line of sight
(Turner et al. 1997; Bassani et al. 1999). In the local universe,
about half of Sy2s are found to be Compton-thick sources with
$N_{\rm H}>10^{24}$ cm $^{-2}$ (Risaliti et al. 1999; Cappi et al. 2006).

 However, up to 50\% Sy2s do not show PBLs in the spectropolarimetric
observations (Tran 2001, 2003; Gu \& Huang 2002).
Several studies indicated that the presence or absence of PBLs depends
on the AGN luminosity, with PBL sources having larger luminosities
(Tran 2001; Lumsden \& Alexander 2001). More specifically, Nicastro et al.
(2003) suggested
that the absence of PBLs corresponds to low values of accretion
rate on to the central black hole. While Tran (2001) proposed the existence
 of a population of AGNs which are intrinsically weak and lack
of BLRs, Lumsden et al. (2001) argued that the detectability of
PBLs is mainly determined by the relative luminosity of the active
core to the host galaxy
(see also Alexander 2001; Gu et al. 2001).

Recently, by focusing on Sy2s with
L$_{\rm [O~III]}$ $>$10$^{41}$ erg s$^{-1}$, Shu et al. (2007,
hereafter Paper I) have shown that,
in additional to the luminosity, the nuclear obscuration also
plays a significant role in the visibility of PBL. In Paper I we found that
Sy2s without PBLs
(hereafter NPBL Sy2s) are more obscured in X-rays than Sy2s with
PBLs ( hereafter PBL Sy2s).
This is in good agreement with the model proposed in Heisler et al. (1997),
who suggested that the absence of PBLs could be attributed
to edge-on line of sight and hidden of electron scattering region (see also 
Taniguchi \& Anabuki 1999).
%Paper I pointed out that the role of obscuration on
%the visibility of PBLs could be revealed only in a luminous Sy2 sample
%where the luminosity dominance is weak enough.

The sample discussed in Paper I includes 27 PBL Sy2s and
15 NPBL Sy2s with L$_{\rm [O~III]}$ $>$10$^{41}$ erg s$^{-1}$ and
X-ray observations available.
%However, we note that there are $\sim$ 30\% (18 out of 60) luminous Sy2s\footnote{Five Sy2s at z$<$0.06
%whose [O {\sc iii}] $\lambda$5007 flux is not available from the
%published literature are excluded from analysis.} at z$<$0.06,
%do not have X-ray observations. 
%The possible bias due to sample
%incompleteness might affect the results presented in Papler I.
% More X-ray observations of Sy2s are required to make the sample X-ray complete
%to reach a robust conclusion.
In this paper we present new released $XMM-Newton$ 0.5 -- 10 keV
spectra of seven Sy2s with optical spectropolarimetric observations,
six of which have L$_{\rm [O~III]}$ $>$10$^{41}$ erg s$^{-1}$.
These
new X-ray observations enlarge our sample of NPBL Sy2s from 15 to 17, and
PBL Sy2s from 27 to 31, providing a chance to verify the results in Paper I.
Section 2 details the $XMM-Newton$ observations
and data analysis. X-ray spectra are analyzed in Section 3. Results are
discussed in Section 4.
Throughout this paper the cosmological parameters 
H$_{0} = 70$ km s$^{-1}$ Mpc$^{-1}$, $\Omega_{m} = 0.27$,
 and $\Omega_{\lambda} = 0.73$ are adopted. \\

%% Authors can use \cite, \citep and \citet for citation.
%% You may also give a citation as 'Michel et al. 1992', and use Table~1 or Fig.~1
%% and so forth. Using \ref and \label for cross-references of Tables/Figures is
%% a good way in adjusting/adding/removing text, tables or figures.

\section{Observations and data analysis}
\label{sect:Obs}

The $XMM-Newton$ observations presented here have been performed between
May 2005 and January 2006 with the EPIC PN (Str$\ddot{\rm u}$der et al. 2001)
detector and MOS (Turner et al. 2001) cameras
operating in full-frame mode.
In this paper, for simplicity only data from the EPIC
PN camera will be discussed.
   Data have been processed using the Science Analysis Software (SAS version
6.5) and have been analyzed using standard software package (FTOOLS 5.0). The
latest calibration files released by the EPIC team have been used.
 Details of the observations with redshift, coordinates (J2000.0),
dates of the observations,
PN net exposures, observation IDs, extinction-corrected
[O {\sc iii}] $\lambda$5007 luminosities, and spectropolarimetric
properties are reported in Table 1.

The event lists produced from the pipeline were filtered to ignore periods
 of high background flaring, by applying fixed thresholds on the single-event,
E$>$10 keV, $\Delta$t=10 s light curves. The thresholds, as well as the
radius of the source circular extraction regions, were optimized to maximize
 the signal-to-noise ratio. The background counts were extracted from
source-free regions on the same chip. Appropriate response and ancillary files
were created using RMFGEN and ARFGEN tasks in the SAS, respectively.
 Spectra were binned in order to have at least 20 counts in each bin
 to ensures the applicability of the $\chi^2$ statistics. We restricted the analysis of the PN data
 in the 0.5 -- 10 keV range and the spectral fitting were performed using XSPEC  version 11.2 software package (Arnaud 1996). The quoted errors on the model
 parameters correspond to a $90\%$ confidence level for one interesting
 parameter ($\Delta\chi^2=2.71$).

\section{X-ray Spectral analysis}
\label{sect:data}

The X-ray spectra of Seyfert 2 galaxies can be approximated by
a two-component continuum plus an Fe K$\alpha$ line
(e.g., Turner et al. 1997). We apply this model, labeled Model 1,
in our fitting procedure, consisting of an absorbed power-law plus
a unabsorbed power-law as the soft X-ray component to
represent the contribution
of scattered emission from the AGN and/or host galaxy. 
This unabsorbed power-law
is either fixed to the value of intrinsic power-law or left free
 to vary in the spectral fits. Our selection then is based on the comparison
 of the $\chi^2$ statistics obtained in the best fits. The possible presence of a narrow
  emission line centered at 6.4 keV originating from neutral iron
has also been checked, and modeled with a single Gaussian line.
All models discussed in this paper include the Galactic absorption
to the line of sight.
The unfolded $XMM-Newton$ PN spectra after background subtraction are shown
in Figure 1 and compared with model spectra.
The best-fit spectral parameters are summarized in Table 2.

\subsection{Notes on Individual Sources}
NGC 513, MCG -3-58-7, NGC 7682:  Model 1 describes the 0.5 -- 10 keV spectrum
 of all three sources well with
acceptable reduced $\chi^2$ ($\sim$1). The absorption-corrected 2 -- 10 keV
luminosities according to the model are 4.9$\times$10$^{42}$ erg s$^{-1}$
for NGC 513, 3.9$\times$10$^{42}$ erg s$^{-1}$ for MCG -3-58-7, and
2.05$\times$10$^{42}$ erg s$^{-1}$ for NGC 7682.

F02581-1136: The $\it XMM-Newton$ 0.5 -- 10 keV spectrum
 fitted by Model 1 yields an
unusual flat hard power-law ($\Gamma=0.33^{+0.43}_{-1.58}$) with
$\chi^2/dof$=29/17. The possibility
that the flat photon index is due to the pile-up effect is excluded.
On the other hand, there have been two possible spectral models for the
observed flat spectrum of Sy2s, the dual absorbed model, and
the reflection
model. We used the PCFABS model to represent the additional absorbed
component with a covering fraction $C_f$. The dual absorbed model
increased $\Gamma$ to $0.79^{+1.79}_{-1.32}$,
with a final $\chi^2$/dof=20/14.
We then adopted the unabsorbed pure Compton reflection component from
neutral matter (PEXRAV model in XSPEC, e.g., Magdziarz \& Zdziarshi 1995).
Comparing with the dual absorbed model,
such model gives an
equally good description of the X-ray spectrum with an associated
$\chi^2$/dof=22/15. We measured a $\Gamma=1.06^{+0.47}_{-1.69}$
and equivalent width (EW) of Fe K$\alpha$ line of 329$^{+338}_{-220}$ eV.
The observed 2 -- 10 keV flux is
9.6$\times$10$^{-13}$ erg s$^{-1}$ cm$^{-2}$, corresponding to
$T$ ratio = 13.7.
However, the resulting Fe K$\alpha$ line EW and $T$ ratio are not expected in
pure reflection spectrum. We note that in the dual absorbed model, the photon
index is poorly constrained. Taking into account the
complexity of spectrum, we finally adopted the dual model
to describe the spectrum with hard photon index fixed to 1.8 ($\chi^2/dof$
= 21/15), and the main conclusions in this paper will not be affected.
The column density of the fully-covering absorber and
partially-covering absorber are 7.6$^{+8.4}_{-3.6}$$\times$10$^{22}$ cm$^{-2}$
 and 5.62$^{+6.0}_{-2.9}$$\times$10$^{23}$ cm$^{-2}$ with a covering
fraction of 0.88$^{+0.07}_{-0.2}$, respectively.
The absorption corrected 2 -- 10 keV luminosity associated with this model
is 8.4$\times$10$^{42}$ erg s$^{-1}$.

 NGC 1144: The $\it XMM-Newton$
 image of this object shows a bright nuclear source.
The best fits of Model 1 to 0.5 -- 10 keV $\it XMM-Newton$ spectrum yields high intrinsic
absorption ($N_{\rm H}$ = 5.62$\times$10$^{23}$ cm$^{-2}$). The EW of Fe 
K$\alpha$ line at 6.4 keV is 252$^{+81}_{-71}$ eV. The observed 2 -- 10 keV 
flux is 3.14$\times$10$^{-12}$ erg s$^{-1}$ cm$^{-2}$, and the
intrinsic 2 -- 10 keV luminosity is 3.14$\times$10$^{43}$ erg s$^{-1}$.
Prieto et al. (2002) reported this object as an unobscured one, and
0.2 -- 2.4 keV flux of 1.1$\times$10$^{-13}$ erg s$^{-1}$ cm$^{-2}$,
on the basis of fits to $ROSAT$/PSPC spectrum. 
We point out that the soft-band
X-ray data is not sufficient to estimate the X-ray absorption
towards nuclear region (see Fig. 1).
The fitted 0.2 -- 2.4 keV flux in $\it XMM-Newton$ spectrum
is 1.0$\times$10$^{-13}$ erg s$^{-1}$ cm$^{-2}$,
 consistent with the result of $ROSAT$/PSPC spectrum.

NGC 6890: This source is the weakest Sy2 in the present sample with
log L$_{\rm [O~III]}$ = 40.86 erg s$^{-1}$ ($<$41 \es).
The $\it XMM-Newton$ observation of this object shows weak nuclear emission.
The spectrum is described by a power-law ( $\Gamma$ = 2.54$^{+0.47}_{-0.43}$)
 but without intrinsic absorption. The fitted 2 -- 10 keV luminosity is
1.0$\times$10$^{40}$ erg s$^{-1}$, consistent with a low luminosity
nuclei. The observed 2 -- 10 keV flux of 6.9$\times$10$^{-14}$
erg s$^{-1}$ cm$^{-2}$, however, gives the $T$ ratio ($\rm F_{2-10~keV}/F_{[O~III]}$) of 0.14, suggesting
heavy obscuration on nuclear region (Bassani et al. 1999,
Guainazzi et al. 2005). Here, we consider it as a Compton-thick one, and
give a a lower limit of 10$^{24}$ cm $^{-2}$ to N$_{\rm H}$,
though longer exposure observations are crucial to shed light on
the nature of absorption in X-rays in this object.

UGC 6100: The $\it XMM-Newton$ observation of UGC 6100 have been strongly affected
by bright background. The poor signal-to-noise prevented us from attempting
detailed spectral fits. Here we give only a very rough, approximate
description of the 0.5 -- 10 keV spectrum in terms of a single power-
law with photon index fixed at 1.8. The resulting 2 -- 10 keV flux is
5.0$\times$10$^{-14}$ erg s$^{-1}$ cm$^{-2}$ and $T$ ratio is 0.05 ($<$0.1),
indicating Compton-thick obscuration (Guainazzi et al. 2005). 
 Note that UGC 6100 has the highest [O {\sc iii}] luminosity in the 
present sample, which is related to strong AGN emission. The observed 
low X-ray flux and thus small $T$ ratio can only be attributed to 
heavily nuclear obscuration. We regard it as a Compton-thick one 
in this paper and give a a lower
limit of 10$^{24}$ cm $^{-2}$ to N$_{\rm H}$.
The observed 2 -- 10 keV luminosity is 1.0$\times$10$^{41}$ erg s$^{-1}$.

\section{Discussion}
%Although it is hard to build an unbiased, complete sample of Sy2s with
% optical spectropolarimetry and X-ray observations to study the relation
% between the presence of PBLs and absorption, the seven targets studied here
% all show evidence of various obscuration level when observed in X-rays.
We find that all four Seyfert 2 galaxies with PBLs are absorbed with 
$N_{\rm H}<10^{24}$ cm$^{-2}$, while two of three Seyfert 2 galaxies 
without PBLs have evidence suggesting
Compton-thick obscuration.
This might indicates that the central region of these Sy2s is viewed differently
 and the lack of PBLs can be ascribed to obscuration effects. On the other
hand, this obscuration might be a signature for the orientation of dusty
torus, shedding some light on the nature of nuclear obscuration matter.

It is instructive to show where these seven Sy2s locate in the plots of
[O {\sc iii}] $\lambda$5007 luminosity versus nuclear obscuration, comparing
with the sample we presented in Paper I.
%compare the X-ray properties of these Sy2s with those
% of other AGNs with spectropolarimetric observations presented by
%Paper I. 
Figure 2 shows the plot of the luminosity of the extinction-corrected 
[O {\sc iii}] $\lambda$5007 emission versus different indicators
of nuclear obscuration ($N_{\rm H}$, $T$ ratio, and Fe K$\alpha$ line EW).
All the 42 sources in Paper I and 7 new sources in this paper are plotted.
 The locus of these targets in the diagram are generally consistent with
 the results of Paper I that at
L$_{\rm [O~{\rm III}]}$ $>$ 10$^{41}$ erg s$^{-1}$, NPBL Sy2s are more
obscured than PBL Sy2s. 
Including these objects (except for NGC 6890), we find the confidence level
in the difference of obscuration between luminous PBL and NPBL Sy2s 
(at L$_{\rm [O~{\rm III}]}$ $>$ 10$^{41}$ erg s$^{-1}$) increases from
%also conducted
%a statistical analysis on the destribution of $N_{\rm H}$, $T$ ratio,
%and Fe K$\alpha$ line EW between NPBL and PBL Sy2s to look for the significant
%level
%of the difference. Kolmogorov-Smirnov (K-S) tests\footnote{When there are
% censored data, we use the survival analysis methods ASURV (Feigelson \&
%Nelson 1985) for statistical analysis.} show that the probability
%for the two samples to be extracted from the same parent population is about
%3.68\% for $N_{\rm H}$, 0.6\% for $T$ ratio, and
%2.59\% for Fe K$\alpha$ line EW,
%corresponding to an improvement of the confidence level of the difference
 92.3\% to 96.3\% for $N_{\rm H}$, 99.1\% to 99.4\% for $T$ ratio,
and 95.3\%
to 97.4\% for Fe K$\alpha$ line EW.
We also plot in Figure 3 $\rm F_{2-10~keV}$ vs.
$\rm F_{[O~III]}$ for Compton-thick
Sy2s with and without PBLs. The data of UGC 6100 and NGC 6890 located in
the diagram follow the correlation for NPBL Sy2s,
 which have smaller $T$ ratio than those Sy2s with PBLs. 
The same conclusion can be seen that the smaller $T$ ratio in
NPBL Sy2s can be explained by heavier nuclear obscuration 
and thus higher inclination of torus.

On the other hand, Nicastro et al. (2003) have argued that the driving
parameter to the absence of PBLs in Sy2s is the accretion rate
(in Eddington units), i. e., PBL Sy2s tend to have Eddington accretion rate
above 10$^{-3}$, while NPBL Sy2s lie at $<$ 10$^{-3}$. Note a contrast 
argument was made by Zhang \& Wang (2006) that NPBL Sy2s tend to have
larger accretion rate, like Narrow Line Seyfert 1 galaxies.
We carry out a direct comparison with the findings
 of the above authors. Making use of $L_{\rm X}$ measured by $\it XMM-Newton$, 
which
is unambiguously related the AGN emission, we estimated the bolometric
luminosity of the seven AGNs, assuming a bolometric correction factor of 10
(e.g., Elvis et al. 1994). For two Compton-thick candidates, we applied the
correction factor of 60 to estimate the intrinsic X-ray luminosity
(Panessa et al. 2006). We obtained published stellar velocity dispersions
from Nelson \& Whittle (1995), Garcia-Rissmann et al. (2005) and $\sigma_\ast
\sim$ 152, 123, 219, and 156 km s$^{-1}$ for NGC 513, NGC 7682,
NGC 1144, and UGC 6100, respectively. For NGC 6890 and F02581-1136,
we used FWHM$_{\rm [O~III]}$ from Whittle (1992), and Heisler et al. (1989),
as a proxy for $\sigma_\ast$, $\sigma_\ast$=FWHM$_{\rm [O~III]}$
/2.35/1.34=78 and 91, respectively (Greene \& Ho 2005). From the values of
the $\sigma_\ast$ and thus the black hole mass, using the $M_{BH} - \sigma_\ast$ correlation (Tremaine et al. 2002), we are now able to estimate the
Eddington ratios $L_{bol}/L_{Edd}$ for the six sources. A comparison with
Figure 1 of Nicastro et al. (2003) shows that all six nuclei have Eddington
ratios well above the threshold of 1.0$\times$10$^{-3}$ to separate AGNs
with PBLs from NPBL sources. The similar large Eddington ratios found for
our PBL and NPBL Sy2s are
consistent with the findings of Bian \& Gu (2007), who suggested that
above a given $L_{bol}/L_{Edd}$ threshold of 10$^{-1.37}$, PBL and NPBL Sy2s
show no difference in their Eddington ratios.

In summary, the $\it XMM-Newton$ X-ray observations of seven bright Sy2s
support the diagram that the absence of PBLs is
associated to the higher obscuration level to the nuclear region. 
%This is also
%supported by the findings of Lumsden et al. (2004), who reported a higher
%detection rate of PBLs in Compton-thin Sy2s. 
The relation between the 
visibility of PBL and the nuclear obscuration can be understood in the 
framework of the unified model
if the scattering region resides very close to the nucleus and its
visibility depends on the viewing angle, as suggested by Heisler et al.
(1997). On the other hand, there is
no evidence showing that in fairly powerful AGNs
($L_{\rm X} \sim 10^{42} - 10^{44}$ erg s$^{-1}$) the lack of PBLs
corresponds to low values of accretion rate onto the central black hole.
%Future, longer X-ray observations would be required to
%ultimately reveal the nature of the extremely obscured AGNs.

\begin{acknowledgements}
The authors thank L. L. Fan and Z. Y. Zheng for helpful discussions. 
Support for this work was provided by
Chinese NSF through NSF10473009/NSF10533050, and
the CAS "Bai Ren" project at University of Science and Technology of China.
\end{acknowledgements}

%
%               one-column-spanning table
%________________________________________ Table 2: Use_of_the routines

%\section{This shows the use of appendix}
%When you compress a *.ps file with gzip.exe in DOS/Windows, you get *.psz.
%Linux's counterpart of the DOS/Windows pkzip/pkunzip or winzip are zip/unzip.

%----------------------------------------------------Table1.
\clearpage
\begin{table}[]
  \caption[]{Summary of the $\it XMM-Newton$ Observations}
  %\fns 
 %\tabcolsep 1.5mm
  \begin{center}\begin{tabular}{lcclcrccc}
    \hline
    \noalign{\smallskip} 
   \hline
         Name     & z    & Coordinates & Obs. date   & 
        PN net exp.  & Sequence  & log(L$_{\rm [O~III]}$) & PBL? \\
          &    & (J2000)  &  & (ks) &  & \ecs &  \\
  \hline\noalign{\smallskip}
NGC 513  & 0.0195 & 01 24 26.85 +33 47 58.0 & 2006 Jan 28 & 11.2 & 0301150401 & 41.14 & Y   \\
NGC 1144 & 0.0289 & 02 55 12.20 -00 11 00.8 & 2006 Jan 28 & 8.3 & 0312190401 & 41.87 & N  \\
NGC 6890 & 0.0081 & 20 18 18.13 -44 48 25.1 & 2005 Sep 29 & 1.26 & 0301151001 & 40.86 & N  \\
NGC 7682 & 0.0171 & 23 29 03.93 +03 32 00.0 & 2005 May 27 & 7.14 & 0301150501 & 41.76 & Y  \\
MCG -3-58-7 & 0.0315 & 22 49 37.15 -19 16 26.4 & 2005 May 09 & 4.03 & 0301150301 & 41.93 & Y  \\
F02581-1136 & 0.0299 & 03 00 30.64 -11 24 56.7 & 2006 Jan 23 & 3.79 & 0301150201 & 41.16 & Y  \\
UGC 6100 & 0.0295 & 11 01 34.00 +45 39 14.2 & 2005 Oct 27 & 2.4 & 0301151101 & 42.28 & N \\
  \noalign{\smallskip}\hline
  \end{tabular}\end{center}
\end{table}

%-----------------------------------------------------Table2
\begin{table}
\centering
\begin{minipage}{100mm}
\caption[]{Best-fitting Spectral Parameters}\end{minipage}\vs

\fns
\begin{tabular}{llccccccll}
  \hline
  \hline\noalign{\smallskip}
       Name &  N$_{\rm H_{Gal}}$ & N$_{\rm H}$  &
        $\Gamma_{\rm HX}$ & $\Gamma_{\rm SX}$ & $f_{\rm s}$(\%) &
        EW(Fe K$\alpha$) & $\chi^2/dof$ & F$_{\rm HX}$ & T \\
    (1) & (2) & (3) & (4) & (5) & (6) & (7) & (8) & (9) & (10)    \\
  \hline\noalign{\smallskip}
NGC 513 & 5.16 & 7.2$\pm0.9$ & 1.66$\pm0.17$ & 2.27$\pm0.27$ & 4.14$\pm0.35$ & 156$\pm$55 & 190/228 & 3.8 & 23.8 \\

NGC 1144 & 6.36 & 56.2$^{+10.2}_{-9.5}$ & 1.95$\pm$0.47 & 2.69$\pm$0.25 & 0.86$\pm0.08$ & 252$^{+81}_{-71}$ & 85/93 & 3.14 & 8.0 \\
NGC 6890 & 3.52 & $>100$ & 2.54$^{+0.47}_{-0.43}$ & $\dots$ & $\dots$  & $\dots$ & 8.8/9 & 0.069  & 0.14 \\
NGC 7682 & 5.16 & 93$^{+73.6}_{-33.6}$ & 2.23$^{+0.49}_{-0.36}$ & =$\Gamma_{\rm HX}$  & 0.93$^{+0.17}_{-0.16}$ & 465$^{+1375}_{-355}$  & 5/11 & 0.3 & 0.35\\

MCG -3-58-7 & 2.53 & 18.4$^{+16.2}_{-9.3}$ & 1.89$^{+1.46}_{-1.08}$ & 3.9$^{+0.46}_{-0.41}$ & 6.2$^{+0}_{-1.1}$ & 46($<$342) & 21/19 & 0.75 & 2.1 \\

F02581-1136$^\ast$ & 4.56 & 56.2$^{+49.6}_{-26.2}$ & 1.8$^{\dag}$ & 4.03$^{+0.49}_{-0.44}$ & 3.9$^{+0.6}_{-0.7}$ & 303$^{+270}_{-213}$ & 21/15 & 0.89 & 12.7 \\
UGC 6100 & 1.19 & $>$100 & 1.8$^{\dag}$ & $\dots$ & $\dots$ & $\dots$ & 12/18 & 0.05 & 0.05 \\
  \noalign{\smallskip}\hline
  \end{tabular}

{{\bf Note:} Col. (1): galaxy name. Col. (2):  Galactic absorption
along the line of sight, in units of 10$^{20}$ cm$^{-2}$.  Col. (3) measured
absorption column density, in units of 10$^{22}$ cm$^{-2}$. Col. (4)
power-law photon index. Col. (5) photon index of the soft power-law
component. Col. (6): scattering fraction of the soft component. Col. (7):
Fe K$\alpha$ line equivalent width  in units of eV. Col. (8): chi-squared
 values and number of degrees of freedom (dof). Col (9): observed 2-10
keV fluxes, in units of 10$^{-12}$ erg s$^{-1}$ cm$^{-2}$. Col (10) T:
the ratio of F$_{\rm 2-10~keV}$/F$_{\rm [O~III]}$. $^{\dag}$ : fixed.
$^\ast$: The X-ray spectrum for this galaxy is fitted by the dual
absorbed model (see the text for details).}
\end{table}
%----------------------------------------------figure1
\clearpage
\begin{figure*}
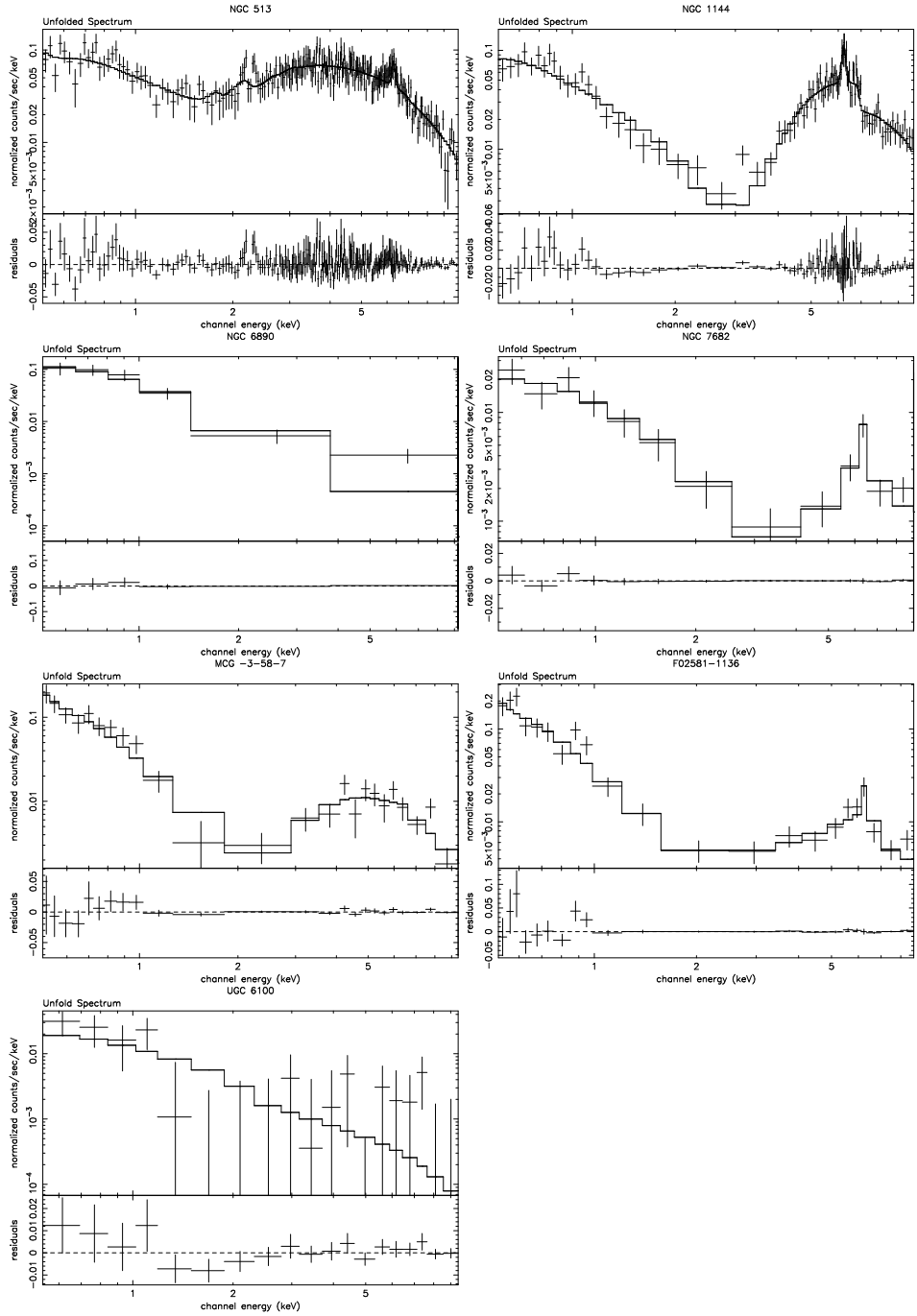

\hbox{
 \includegraphics[width=45mm, angle=-90]{fig1a.eps}
 \includegraphics[width=45mm, angle=-90]{fig1b.eps}
}
\hbox{
 \includegraphics[width=45mm, angle=-90]{fig1c.eps}
 \includegraphics[width=45mm, angle=-90]{fig1d.eps}
}
\hbox{
 \includegraphics[width=45mm, angle=-90]{fig1e.eps}
 \includegraphics[width=45mm, angle=-90]{fig1f.eps} }
\hbox{
 \includegraphics[width=45mm, angle=-90]{fig1g.eps}
}

\caption{ $XMM-Newton$ PN spectra of seven Seyfert 2 galaxies
in our sample. The top panel shows the data and model, and the
data/model ratio is shown in the bottom panel.
\label{fig1}}
\end{figure*}
%-------------------------------------------------figure2
\clearpage
\begin{figure}
\begin{center}
%\epsscale{2.5}{2.5}
\psfig{figure=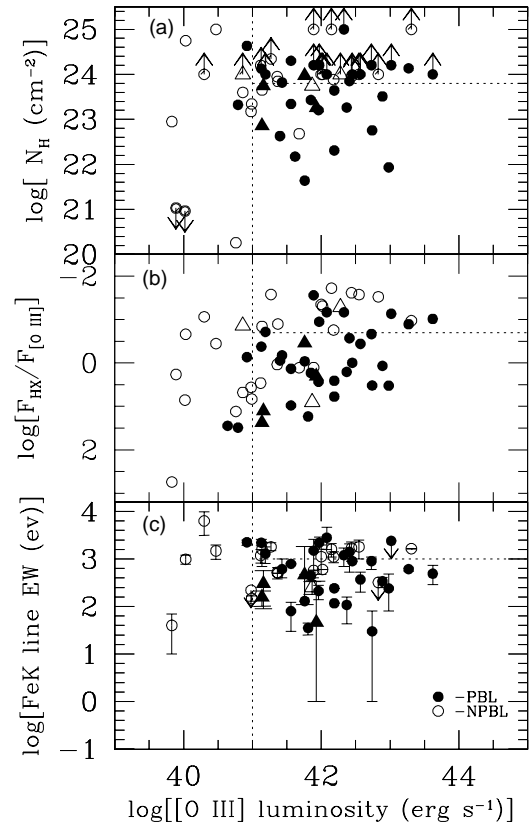,width=120mm}
\caption{Plot of [O {\sc iii}] $\lambda$5007 luminosity vs. different
 obscuration indicators for PBL Sy2s (filled symbols) and
NPBL Sy2s (open symbols). (a) plots [O {\sc iii}] $\lambda$5007
luminosity against $N_{\rm H}$, (b) $T$ ratio,
and (c) Fe K$\alpha$ line EW. The filled circles denote
 Sy2s in Paper I and the triangles represent objects in present sample.
}
\end{center}
\end{figure}
%-------------------------------------------------figure3
\begin{figure}
\centering
\includegraphics[width=90mm]{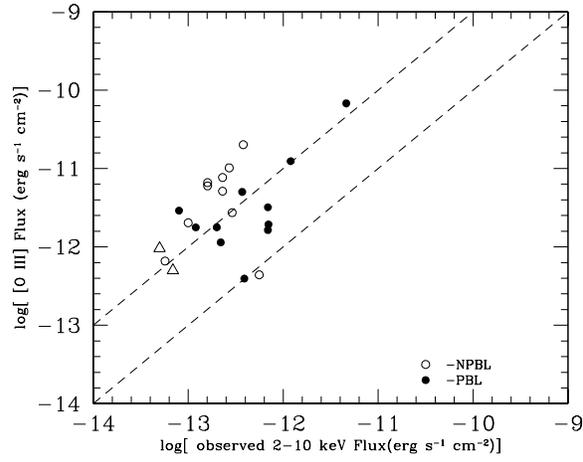}
\vspace{-5mm}

\caption{Observed [O {\sc iii}] flux (extinction-corrected) vs.
X-ray (2 -- 10 keV) flux for Compton-thick Sy2s (with $N_{\rm H}$ $>$ 10$^{24}$
cm$^{-2}$). The dashed lines represent $F_{\rm [O~III]}$ =
10 $F_{\rm 2-10keV}$ (upper) and $F_{\rm [O~III]}$ = $F_{\rm 2-10keV}$ (lower).
Symbols have the same coding as in Figure 2.
\label{fig3}}
\end{figure}

\label{lastpage}

\end{document}